\documentclass{article}
\usepackage{graphicx}

\title{Numerical study of subdiffusion equation}

\author{Katarzyna D. Lewandowska$^1$, Tadeusz Koszto{\l}owicz$^2$}

\date{\footnotesize{$^1$ Department of Physics and Biophysics, Medical University of
         Gda\'nsk,\\ ul. D\c{e}binki 1, 80-211 Gda\'nsk, Poland, \\
         e-mail: kale@amg.gda.pl}\\
        $^2$Institute of Physics, \'Swi\c{e}tokrzyska Academy,\\
         ul. \'Swi\c{e}tokrzyska 15, 25-406 Kielce, Poland, \\
         e-mail: tkoszt@pu.kielce.pl}

\begin{document}
\maketitle

\begin{abstract}
We present a numerical procedure of solving the subdiffusion
equation with Caputo fractional time derivative. On the basis of
few examples we show that the subdiffusion is a 'long time memory'
process and the short memory principle should not be used in this
case.
\end{abstract}

\date{PACS numbers: 02.50.Ey, 05.10.-a, 02.60.Cb}

\section{Introduction}

The subdiffusion equation is of fractional order with respect to
the time variable. Unfortunately, the exact solutions are known
only for relatively simple systems, similarly to the normal
diffusion case. In more complicated situations such as a system
with subdiffusion coefficient depending on concentration,
inhomogenous fractional subdiffusive system or
subdiffusion-reaction system, one needs a numerical procedure to
solve of the equation. Usually the subdiffusion equation contains
the Riemann-Liouville fractional time derivative of the order
$1-\alpha$ ($\alpha$ denotes here the subdiffusion parameter),
which is not convenient for physical interpretation of initial
conditions. To get the subdiffusion equation with initial
conditions which have simple interpretation, one can use the
subdiffusion equation with the Caputo fractional time derivative
of the order $\alpha$. As far as we know, there is a numerical
method to solve the subdiffusion equation with the
Riemann-Liouville fractional time derivative \cite{y}. The
equation with Caputo derivative has been numerically studied only
within the time fractional discrete random walk \cite{gmmp}.

Subdiffusion is a process with the time memory. There arises a
practical problem with the memory length, which extends to
$-\infty$. To omit the difficulty Podlubny \cite{p,pdk} postulated
to apply the short memory principle which assumes that the
relatively small memory length is sufficient to obtain
satisfactorily accuracy of numerical solutions for sufficiently
long times. However, as shown here, the method produces
significant differences between the numerical and exact solutions.
In this paper we present a numerical procedure of solving the
subdiffusion equation with Caputo derivative, which is based on
the fractional difference approach, and we briefly study the
efficiency of the short memory principle for this case.

\section{Subdiffusion equation}

The transport process is described by the subdiffusion equation
\cite{mk}
\begin{equation}\label{eqRL}
\frac{\partial C(x,t)}{\partial
t}=D_{\alpha}\frac{^{RL}\partial^{1-\alpha}}{\partial
t^{1-\alpha}}\frac{\partial^{2}C(x,t)}{\partial x^{2}},
\end{equation}
where $0<\alpha<1$, $C(x,t)$ denotes the concentration of
transported substance, the Riemann-Liouville fractional time
derivative is defined for $\alpha>0$ as
\begin{equation}\label{rl}
\frac{^{RL}\partial^{\alpha}f(t)}{\partial
t^{\alpha}}=\frac{1}{\Gamma(n-\alpha)}\frac{\partial^{n}}{\partial
t^{n}}\int_{0}^{t}dt'\frac{f(t')}{(t-t')^{1+\alpha-n}},
\end{equation}
where the integer number $n$ fulfills the relation $n-1<\alpha\leq
n$.

The presence of time derivatives on both sides of Eq.~(\ref{eqRL})
is not convenient for numerical calculations. To simplify the
numerical procedure we rewrite the Eq.~(\ref{eqRL}) in the form
\begin{equation}\label{eqC}
\frac{^{C}\partial^{\alpha}C(x,t)}{\partial
t^{\alpha}}=D_{\alpha}\frac{\partial^{2}C(x,t)}{\partial x^{2}},
\end{equation}
where the fractional time derivative on the left-hand side of
Eq.~(\ref{eqC}) is now Caputo derivative defined by the relation
\begin{equation}\label{c}
\frac{^{C}\partial^{\alpha}f(t)}{\partial
t^{\alpha}}=\frac{1}{\Gamma(\alpha-n)}\int_{0}^{t}dt'
\frac{f^{(n)}(t')}{(t-t')^{1+\alpha-n}},
\end{equation}
$f^{(n)}$ denotes the derivative of natural order $n$.

We note that the Laplace transform of the Caputo fractional
derivative involves values of the function $f$ and its derivatives
of natural order at $t=0$ while Laplace transform of the
Riemann-Liouville fractional derivative includes the fractional
derivatives of $f$ at $t=0$. Thus, the physical interpretation of
initial conditions in the former case is clear in contrast to the
latter one \cite{p}. So, it is more convenient to set the initial
conditions for the equation with Caputo derivative than for the
equation with Riemann-Liouville one.

\section{Numerical procedure}

\subsection{Fractional derivatives}

To numerically solve the normal diffusion equation one usually
substitutes the time derivative by the backward difference
$\frac{\partial f(t)}{\partial t}\simeq \frac{f(t)-f(t-\Delta
t)}{\Delta t}$. In the presented procedure we proceed in a similar
way. For that purpose we use the Gr\"{u}nwald-Letnikow fractional
derivative, which is defined as a limit of a fractional-order
backward difference \cite{p}
    \begin{equation}\label{gl}
\frac{^{GL}\partial^{\alpha}f(t)}{\partial
t^{\alpha}}=\lim_{\Delta t\rightarrow 0}(\Delta
t)^{-\alpha}\sum_{r=0}^{[\frac{t}{\Delta t}]}(-1)^{r} \left(
\begin{array}{c}
\alpha\\
r
\end{array} \right)f(t-r\Delta t),
    \end{equation}
where $\alpha>0$, $[x]$ means the integer part of $x$ and
\begin{displaymath}
\left( \begin{array}{c}
\alpha\\
r
\end{array} \right)=\frac{\Gamma(\alpha+1)}{r!\Gamma(\alpha-r+1)}
=\frac{\alpha(\alpha-1)(\alpha-2)\cdot\ldots\cdot
[\alpha-(r-1)]}{1\cdot 2\cdot 3\cdot\ldots\cdot r}.
\end{displaymath}

When the function $f(t)$ of positive argument has continuous
derivatives of the integer order $0,1,\ldots,n$, the
Riemann-Liouville definition~(\ref{rl}) is equivalent to the
Gr\"{u}nwald-Letnikow one \cite{p}. So, we can take
\begin{equation}\label{glrl}
\frac{^{RL}\partial^{\alpha}f(t)}{\partial
t^{\alpha}}=\frac{^{GL}\partial^{\alpha}f(t)}{\partial
t^{\alpha}}.
\end{equation}
The relation between Riemann-Liouville and Caputo derivatives is
more complicated and reads as \cite{p}
\begin{equation}\label{rlc}
\frac{^{RL}\partial^{\alpha}f(t)}{\partial
t^{\alpha}}=\frac{^{C}\partial^{\alpha}f(t)}{\partial
t^{\alpha}}+\sum_{k=0}^{n-1}\Phi_{k-\alpha+1}(t)f^{(k)}(0),
\end{equation}
where
\begin{equation}\label{phi}
\Phi_{q+1}(t)=\left\{
\begin{array}{cc}
\frac{t^{q}}{\Gamma(q+1)}&t>0\\
0&t\leq 0
\end{array}
\right. .
\end{equation}
From Eqs. (\ref{gl})-(\ref{phi}) we can express the Caputo
fractional derivative in terms of the fractional-order backward
difference \cite{p}
\begin{equation}\label{glc}
\frac{^{C}\partial^{\alpha}f(t)}{\partial t^{\alpha}}=\lim_{\Delta
t\rightarrow 0}(\Delta t)^{-\alpha}\sum_{r=0}^{[\frac{t}{\Delta
t}]}(-1)^{r}\left(
\begin{array}{c}
\alpha\\
r
\end{array}
\right)f(t-r\Delta t)-\frac{1}{t^{\alpha}\Gamma(1-\alpha)}f(0).
\end{equation}

\subsection{Algorithm}

The standard way to approximate of the fractional derivative,
which is useful for numerical calculations, is to omit the limit
in Eq.~(\ref{glc}) and to change the infinite series occurring in
(\ref{glc}) to the finite one
\begin{equation}\label{glca}
\frac{^{C}\partial^{\alpha}f(t)}{\partial t^{\alpha}}\simeq(\Delta
t)^{-\alpha}\sum_{r=0}^{L}(-1)^{r} \left(
\begin{array}{c}
\alpha\\
r
\end{array}
\right)f(t-r\Delta t)-\frac{1}{t^{\alpha}\Gamma(1-\alpha)}f(0),
\end{equation}
where arbitrary chosen parameter $L$ is called the memory length.
Substituting Eq.~(\ref{glca}) to Eq.~(\ref{eqC}) and using the
following approximation of the second order derivative
\begin{equation}
\frac{\partial^{2}f(x)}{\partial x^{2}}\simeq \frac{f(x+\Delta
x)-2f(x)+f(x-\Delta x)}{(\Delta x)^{2}},
\end{equation}
after simple calculation we obtain
\begin{eqnarray}\label{alg}
\nonumber
C(x,t)=-\sum_{r=1}^{L}(-1)^{r}\frac{\alpha(\alpha-1)(\alpha-2)
\cdot\ldots\cdot[\alpha-(r-1)]}{1\cdot 2\cdot 3\cdot\ldots\cdot r
}C(x,t-r\Delta t)\\+\frac{1}{t^{\alpha}\Gamma(1-\alpha)}C(x,0)
+D_{\alpha}\frac{(\Delta t)^{\alpha}}{(\Delta x)^{2}}[C(x+\Delta
x,t-\Delta t)\\ \nonumber -2C(x,t-\Delta t)+C(x-\Delta x,t-\Delta
t)].
\end{eqnarray}
There arise a problem with the choice of initial conditions. We
choose the initial conditions only for $t=0$ assuming that the
concentration given for earlier moments $t<0$ does not influence
the process for $t>0$. This assumption is in agreement with the
procedure of solving the equations with Caputo fractional
derivative where the initial conditions are determined only at
$t=0$ for the derivatives of natural order $f^{(n)}(t)|_{t=0}$,
$n=0,1,\ldots,[\alpha]$ (here $f^{(0)}\equiv f$). In our
considerations we have $0<\alpha<1$, hence it is enough to set
$C(x,0)$ as the initial condition.

Starting with the initial condition $C(x,0)$ we will find the time
iterations $C(x,t_s \Delta t)$ for $t_s=1,2,...,t_{s,max}$. When
the number of time steps $t_{s}$ is less then the memory length
$L$ then we put $L=t_{s}$ in the series occurring in Eq.
(\ref{alg}), otherwise the memory length is equal to $L$.

\section{Numerical results}

To test the numerical procedure we are going to compare the
numerical solutions of the subdiffusion equation with the exact
analytical ones. For that purpose we choose the homogenous system
with the initial concentration
\begin{equation}\label{icon}
C(x,0)=\left\{ \begin{array}{cc}
C_{0} & x<0\\
0 & x\geq 0
\end{array} \right. .
\end{equation}
The solution of the subdiffusion equation~(\ref{eqC}) with the
initial condition~(\ref{icon}) is following \cite{kdm}
\begin{equation}\label{sol}
C(x,t)=\left\{
\begin{array}{lc}
C_{0}-\frac{C_{0}}{\alpha}H^{1 0}_{1 1}\left(
\frac{(-x)^{2/\alpha}}{D_{\alpha}^{1/\alpha}t} \left|
    \begin{array}{cc}
             1 & 1 \\
             0 & 2/\alpha
    \end{array} \right. \right) &x<0 \\ & \\
\frac{C_{0}}{\alpha}H^{1 0}_{1
1}\left(\frac{x^{2/\alpha}}{D_{\alpha}^{1/\alpha}t} \left|
    \begin{array}{cc}
             1 & 1 \\
             0 & 2/\alpha
    \end{array} \right. \right) &x\geq 0
\end{array} \right. ,
\end{equation}
where $H$ denotes the Fox function, which can be expressed by the
series \cite{k1}
    \begin{equation}
    \label{h}
H^{1 0}_{1 1}\left( \frac{x^{2/\alpha}}{D_{\alpha}^{1/\alpha}t}
\left|
    \begin{array}{cc}
             1 & 1 \\
             0 & 2/\alpha
    \end{array} \right. \right)
=\frac{2}{\alpha}\sum_{k=0}^{\infty}\frac{1}{k!\Gamma \left
(1-k\alpha/2 \right )}\left
(-\frac{x}{\sqrt{D_\alpha}t^{\alpha/2}}\right )^{k}\;.
    \end{equation}

\begin{figure}[h!]\label{t100}
\centering
\includegraphics[height=10cm]{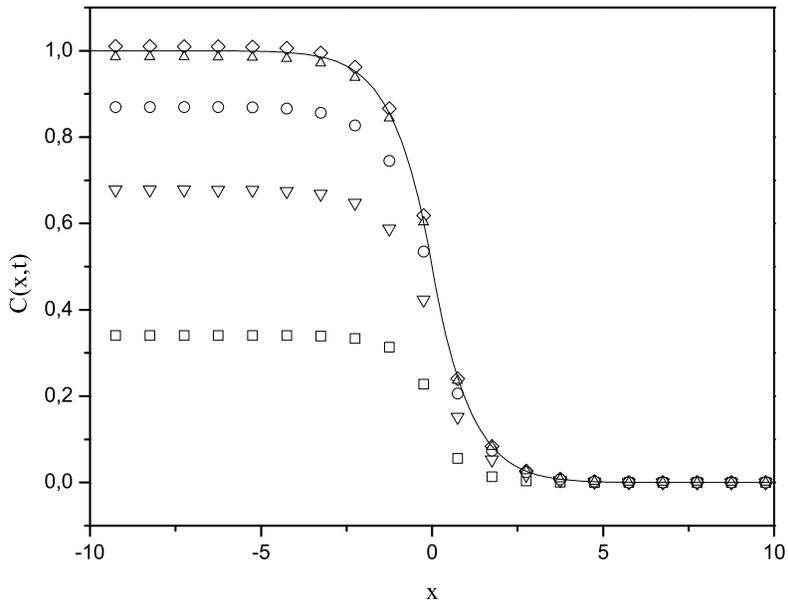}
\caption{The concentration profiles $C(x,t)$ calculated for
$\alpha=0.5$, $D_{\alpha}=0.25$, $t_{s,max}=100$ and with
different memory length $L=100(\diamondsuit)$, $80(\triangle)$,
$50(\circ)$, $30(\nabla)$, and $10$; continuous line represents
the exact analytical solution.}
\end{figure}

\begin{figure}[h!]\label{bt50}
\centering
\includegraphics[height=10cm]{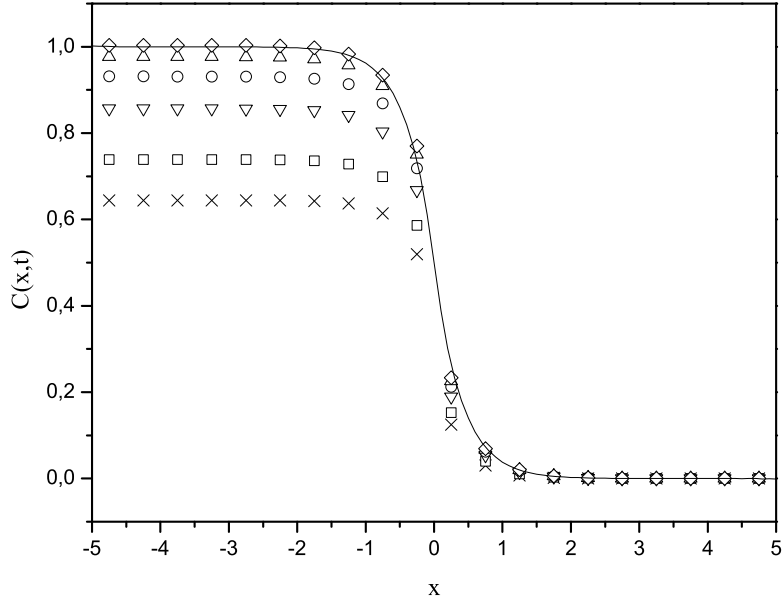}
\caption{The concentration profiles obtained for $\alpha=0.2$,
$D_{\alpha}=0.2$, $t_{s,max}=50$ and with memory length
$L=50(\diamondsuit)$, $40(\triangle)$, $30(\circ)$, $20(\nabla)$,
$10$, and $5(\times)$.}
\end{figure}

\begin{figure}[h!]\label{ct50}
\centering
\includegraphics[height=10cm]{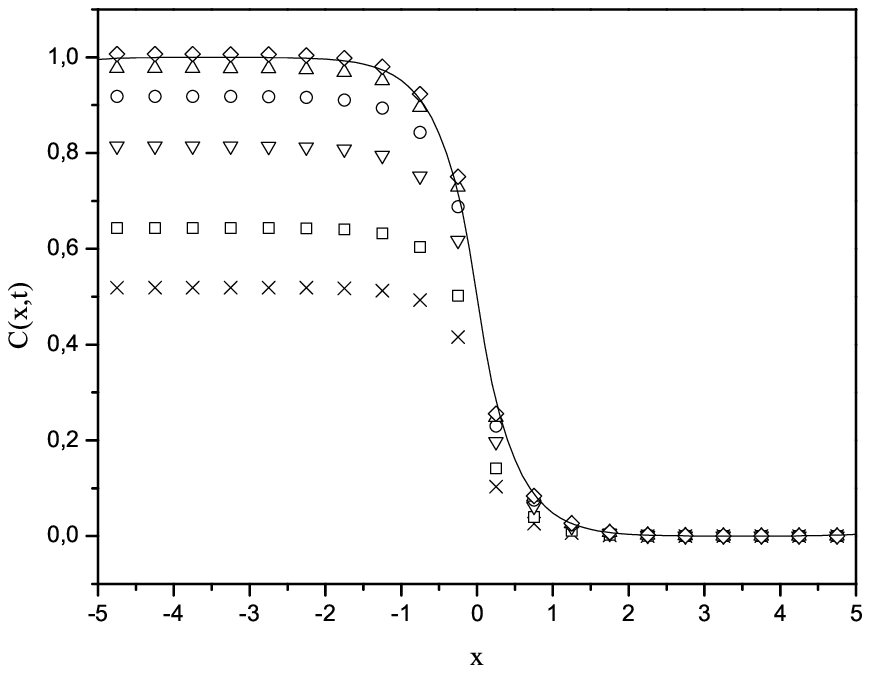}
\caption{The concentration profiles calculated for $\alpha=0.3$,
$D_{\alpha}=0.1$, time $t_{s,max}=50$ and with memory length
$L=50(\diamondsuit)$, $40(\triangle)$, $30(\circ)$, $20(\nabla)$,
$10$, and $5(\times)$.}
\end{figure}

The results of numerical calculations and the analytical solutions
are shown in the plots. In Figures 1-3 we present the numerical
solutions of the subdiffusion equation for different values of
$t$, $\alpha$, and $D_{\alpha}$. In each case we present the plot
of analytical solution (continuous line) and numerical solutions
calculated for different memory length $L$ (symbols without line).
The time and the memory length are given as the number of all time
steps $t_{s,max}$, which corresponds to the 'real time' $t$ by the
relation $t=\Delta t\cdot t_{s,max}$. In all cases we take
$C_0=1$, $\Delta t=0.1$ and $\Delta x=0.5$ (all quantities are
given in arbitrary units); to calculate the analytical solutions
(\ref{sol}) we took 100 first terms in the series occurring in
(\ref{h}). We can see that the memory length determines the
accuracy of numerical solutions.

\section{Final remarks}

We have presented the procedure to numerically solve the
subdiffusion equation with Caputo fractional time derivative. The
choice of the equation in such a form is not accidental since the
interpretation of the initial condition in this case is simpler
than in the equation with Riemann-Liouville derivative. In all
considered cases the numerical solutions coincide with the
analytical ones. In the studies \cite{p,pdk} the 'short memory
principle' was postulated. According to this principle, the
fractional derivative is approximated by the fractional derivative
with moving lower limit $t-L$, where $L$ is the 'memory length'.
The examples presented in \cite{p} suggest that the $L=50$ time
steps gives a good approximation for times of the order of
$t_s\sim 100$ time steps. However, the results presented here show
that this memory length is not sufficient for the subdiffusion
case. Our analysis demonstrate that the memory length should be
longer than about 80 per cent of the value of time variable.

\section*{Acknowledgements}

The authors wish to express his thanks to Stanis{\l}aw
Mr\'owczy\'nski for fruitful discussions and critical comments on
the manuscript. This paper was supported by Polish Ministry of
Education and Science under Grant No. 1 P03B 136 30.


\begin{thebibliography}{33}

\bibitem{mk} R. Metzler, J. Klafter, \textit{Phys. Rep.} \textbf{339}, 1
(2000); \textit{J. Phys.} \textbf{A37}, R161 (2004).
\bibitem{y} S.B. Yuste, L. Acedo, \textit{SIAM J. Numer. Anal.}
\textbf{42}, 1862 (2005) (and references therein).
\bibitem{gmmp} R. Gorenflo, F. Mainardi, D. Moretti, P.
Paradisi, \textit{Nonlin. Dyn.} \textbf{29}, 129 (2002).
\bibitem{p} I. Podlubny, \textit{Fractional differential equations}; Academic
Press, San Diego 1999.
\bibitem{pdk} I. Podlubny, L. Dorcak, I. Kostial,
\textit{Proc. 36th IEEE CDC}, San Diego 1997, p. 4985 .
\bibitem{kdm} T. Koszto{\l}owicz, K. Dworecki, S.
Mr\'{o}wczy\'{n}ski, \textit{Phys. Rev.} \textbf{E71}, 041105
(2005).
\bibitem{k1} T. Koszto{\l}owicz, \textit{J. Phys. A: Math. Gen.} {\bf 37}, 10779 (2004).

\end{thebibliography}
\end{document}